\documentclass[aip,jcp,numerical,preprint]{revtex4-1}
\usepackage{color}
\usepackage{graphicx}
\usepackage{threeparttable}
\usepackage{longtable}
\usepackage{enumerate}
\usepackage{amsfonts,amssymb}
\usepackage{amsmath}
\usepackage{stmaryrd}
\usepackage{dcolumn}
\usepackage{bm}
\usepackage{bbm}
\usepackage{algpseudocode}
\usepackage{algorithm}
\usepackage[all,cmtip]{xy}
\usepackage{url}

\bibpunct{}{}{,}{s}{}{\textsuperscript{,}}  % upper cite

\newcommand{\ii}{\mathbbm{i}}

\newcommand{\spn}{\mathrm{span}}
\newcommand{\hc}{\mathrm{h.c.}}
\newcommand{\rR}{\mathrm{R}}
\newcommand{\rI}{\mathrm{I}}

\begin{document}
%%% ----------------------------------------------------------------------

\title{
Time-reversal symmetry adaptation in relativistic density matrix renormalization group algorithm
}
\author{Zhendong Li}\email{zhendongli2008@gmail.com}
\affiliation{Key Laboratory of Theoretical and Computational Photochemistry, Ministry of Education, College of Chemistry, Beijing Normal University, Beijing 100875, China}
\date{\today}

\begin{abstract}
In the nonrelativistic Schr\"{o}dinger equation, the total spin $S$ and spin projection $M$ are good quantum numbers.
In contrast, spin symmetry is lost in the presence of spin-dependent interactions such as spin-orbit couplings in relativistic Hamiltonians.
Previous implementations of relativistic density matrix renormalization group algorithm (R-DMRG) only employing particle number symmetry
are much more expensive than nonrelativistic DMRG. Besides, artificial breaking of Kramers degeneracy can happen
in the treatment of systems with odd number of electrons. To overcome these issues, we introduce time-reversal symmetry adaptation for R-DMRG.
Since the time-reversal operator is antiunitary, this cannot be simply achieved in the usual way.
We define a time-reversal symmetry-adapted renormalized
basis and present strategies to maintain the structure of basis functions
during the sweep optimization. With time-reversal symmetry
adaptation, only half of the renormalized operators are needed and the computational
costs of Hamiltonian-wavefunction multiplication
and renormalization are reduced by half. The present construction of time-reversal
symmetry-adapted basis also directly applies to other tensor network states without loops.
\end{abstract}
\maketitle

\section{Introduction}
Relativistic quantum mechanics are more accurate description of the real world
than the nonrelativistic quantum mechanics\cite{dyall2007introduction}. Phosphorescence, intersystem
crossings, and zero-field splittings cannot be described by the Schr{\"o}dinger equation due to the lack
of spin-dependent interactions such as spin-orbit couplings (SOC) and spin-spin interactions.
Unfortunately, it is known that relativistic calculations are much more expensive that nonrelativistic
calculations either at the mean-field or correlated level.
Depending on the level of theories, this can be attributed to several factors, such as the presence of negative energy states in the
four-component Dirac equation, the loss of spin symmetry, and the use of complex algebra, etc.
At the correlated level, adopting the no-pair approximation in four-component theories
or using a two-component relativistic Hamiltonian such as
the exact two-component (X2C) Hamiltonian\cite{liu2010ideas,saue2011relativistic,peng2012exact}
from the start will make the dimension of the one-electron basis identical to that in
the nonrelativistic unrestricted case\cite{fleig2012invited}.
The loss of spin symmetry is a more challenging issue.
Since the time-reversal operator $\mathcal{T}$
commutes with relativistic Hamiltonians $\hat{H}$
in the absence of magnetic field, the time-reversal symmetry can be used to ameliorate the
situation. However, its adaptation in correlated methods is nontrivial
\cite{visscher1995kramers,fleig2008time,tu2011symmetry,fleig2001generalized,zhang2022soici}.

In this work, we consider the time-reversal symmetry adaptation in
relativistic density matrix renormalization group (R-DMRG) algorithm.
The DMRG algorithm\cite{white1992density} has become a powerful tool
for treating strongly correlated molecules\cite{white1999ab,chan2002highly,
chan2011density,Reiher2011,Sebastian2014,Legeza2015,Yanai2015,olivares-amaya_ab-initio_2015,baiardi2020density,ma2022post}.
Thus, it will be of great interest to apply it to challenging systems
involving heavy elements. In fact, including scalar relativistic effects into DMRG
was carried out long time ago\cite{moritz2005relativistic,nguyen2014scalar}, and
state interaction schemes for treating SOC
based on matrix product states (MPS) obtained
from spin-free DMRG calculations
has been put forward\cite{roemelt2015spin,sayfutyarova2016state,knecht2016nonorthogonal}.
However, including SOC variationally within DMRG\cite{knecht2014communication,battaglia2018efficient,zhai2022comparison,hoyer2022relativistic}
has only been achieved without time-reversal symmetry adaptation.
Such R-DMRG implementation is much more expensive than nonrelativistic non-spin-adapted DMRG.
While in the nonrelativistic case, a renormalized state can be
labeled by $|N,M\rangle$ (point group symmetry is not discussed here), where $N$ is the particle number and $M$ is the spin projection,
only $N$ is a good quantum number in the relativistic case.
Another problem for lacking time-reversal adaptation is that
artificial breaking of Kramers degeneracy can happen
in the treatment of systems with odd number of electrons.
These drawbacks motivate us to introduce time-reversal symmetry adaptation for R-DMRG.

Conceptually, the fundamental difficulty of time-reversal symmetry adaptation is that
the time-reversal operator $\mathcal{T}$ is an anti-unitary operator\cite{wigner1959group},
unlike other symmetry operations employed in DMRG.
In simple words, it cannot be simply achieved by associating a 'quantum number' for an
irreducible representation to renormalized states\cite{mcculloch2002non,zgid2008spin,singh2010tensor,singh2011tensor,singh2012tensor,
weichselbaum2012non,sharma2012spin,keller2016spin}.
In principle, one can add time-reversal operation into a symmetry group
to form an enlarged group (called magnetic group\cite{bradley2010mathematical}),
and use Wigner's corepresentation theory\cite{wigner1959group}, which
is a generalization of the standard representation theory for groups of unitary operators
to groups including anti-unitary elements. Without going into such mathematical complication, we will show that introducing
a concept of time-reversal symmetry-adapted renormalized basis is already sufficient
for our purpose. The proposed usage of time-reversal symmetry-adapted basis
also directly applies to other tensor network states (TNS) without loops,
such as the general tree TNS\cite{shi2006classical,murg2010simulating,
nakatani2013efficient,murg2015tree} or the simpler comb TNS\cite{chepiga2019comb,li2021expressibility}.

The remainder of this paper is organized as follows. In Sec. II, we
introduce the definition of time-reversal symmetry-adapted basis. In Sec. III,
we present strategies to maintain such structure during the sweep optimization
in DMRG, and demonstrate that a reduction of the computational costs of matrix-vector multiplication and renormalization
by half can be achieved by using time-reversal symmetry.
Conclusion and outlook is given in the last section.

\section{Time reversal symmetry-adapted basis}
\subsection{Definition}
To utilize the time-reversal symmetry, we introduce the following
\emph{time-reversal symmetry-adapted orthonormal basis}
for a subspace $V_l$ of the Fock space
\begin{eqnarray}
V_l &=& V_l^{\mathrm{e}}\oplus V_l^{\mathrm{o}}, \nonumber\\
V_l^{\mathrm{e}} &=& \spn\{|l_{\mathrm{e}}\rangle\}, \quad \mathcal{T}|l_{\mathrm{e}}\rangle = |l_{\mathrm{e}}\rangle,\nonumber\\
V_l^{\mathrm{o}} &=& \spn\{|l_{\mathrm{o}}\rangle,\;|l_{\bar{\mathrm{o}}}\rangle\}, \quad |l_{\bar{\mathrm{o}}}\rangle \triangleq \mathcal{T}|l_{\mathrm{o}}\rangle, \label{eq:VlRel}
\end{eqnarray}
where the superscript/subscript e (or o) represents even (or odd) number of electrons.
Here, the notation $V_l^{\mathrm{e}}=\spn\{|l_{\mathrm{e}}\rangle\}$ needs to be understood as $V_l^{\mathrm{e}}=\spn\{|l_{\mathrm{e}}^1\rangle,
|l_{\mathrm{e}}^2\rangle,\cdots,|l_{\mathrm{e}}^m\rangle\}$, and we only show one of the basis functions
for brevity. For later convenience, we will refer to the structure of basis for $V_l^{\mathrm{e}}$ as time-reversal \emph{invariant},
and that for $V_l^{\mathrm{o}}$ as time-reversal \emph{paired}. Note that the orthogonality between $|l_{\mathrm{o}}\rangle$ and its time-reversal partner $|l_{\bar{\mathrm{o}}}\rangle$
is automatically guaranteed by the Kramers' theorem.
An arbitrary basis for $V_l$ does not necessarily takes this form \eqref{eq:VlRel}.
However, as long as $V_l$ is invariant under the action of $\mathcal{T}$, we can
always construct basis functions with such structure based on the following observations:

(i) For the even-electron case, an arbitrary state
$|l_{\mathrm{e}}\rangle$ and its time-reversal partner $|l_{\bar{\mathrm{e}}}\rangle=\mathcal{T}|l_{\mathrm{e}}\rangle$ is not necessarily orthogonal,
because the time-reversal operation does not impose any constraint on the overlap $\langle l_{\mathrm{e}}|l_{\bar{\mathrm{e}}}\rangle$.
Suppose $|l_{\mathrm{e}}\rangle$ is normalized, then $|\langle l_{\mathrm{e}}|l_{\bar{\mathrm{e}}}\rangle|\le 1$ by the
Cauchy-Schwarz inequality. There can be two cases:

\begin{enumerate}
\item If $|\langle l_{\mathrm{e}}|l_{\bar{\mathrm{e}}}\rangle|=1$, which means that these two states are parallel and
simply differ by a phase $|l_{\bar{\mathrm{e}}}\rangle = e^{\ii\phi}|l_{\mathrm{e}}\rangle$,
then the new state $|\mathcal{L}_{\mathrm{e}}\rangle \triangleq e^{\ii\phi/2}|l_{\mathrm{e}}\rangle$
will satisfy the condition in Eq. \eqref{eq:VlRel}, i.e.,
$|\mathcal{L}_{\bar{\mathrm{e}}}\rangle \triangleq \mathcal{T}|\mathcal{L}_{\mathrm{e}}\rangle = |\mathcal{L}_{\mathrm{e}}\rangle$.
This derivation also shows that in this case the eigenvalue of $\mathcal{T}$ is arbitrary, since
$\mathcal{T}e^{\ii\theta}|l_{\mathrm{e}}\rangle = e^{-\ii\theta}|l_{\bar{\mathrm{e}}}\rangle = e^{-\ii(2\theta-\phi)}
(e^{\ii\theta}|l_{\mathrm{e}}\rangle)$ for an arbitrary $\theta$. Our choice $\theta=\phi/2$
will make the later discussion of matrix representation of operators very compact.

\item If $|\langle l_{\mathrm{e}}|l_{\bar{\mathrm{e}}}\rangle|<1$, then the linear combinations
\begin{eqnarray}
\left\{\begin{array}{ccc}
|\mathcal{L}_{\mathrm{e}}^+\rangle &=& \frac{1}{\sqrt{2}}(|l_{\mathrm{e}}\rangle + |l_{\bar{\mathrm{e}}}\rangle), \\
|\mathcal{L}_{\mathrm{e}}^-\rangle &=& \frac{\ii}{\sqrt{2}}(|l_{\mathrm{e}}\rangle - |l_{\bar{\mathrm{e}}}\rangle),
\end{array}\right.
\end{eqnarray}
yield two linear independent functions satisfying Eq. \eqref{eq:VlRel}.
They are not orthonormal as the overlap metric depends on the overlap between $|l_{\mathrm{e}}\rangle$
and $|l_{\bar{\mathrm{e}}}\rangle$,
\begin{eqnarray}
\begin{bmatrix}
 \langle \mathcal{L}_{\mathrm{e}}^+ |\mathcal{L}_{\mathrm{e}}^+\rangle &  \langle \mathcal{L}_{\mathrm{e}}^+ |\mathcal{L}_{\mathrm{e}}^-\rangle \\
 \langle \mathcal{L}_{\mathrm{e}}^- |\mathcal{L}_{\mathrm{e}}^+\rangle &  \langle \mathcal{L}_{\mathrm{e}}^- |\mathcal{L}_{\mathrm{e}}^-\rangle \\
\end{bmatrix}
=
\begin{bmatrix}
1+\Re\langle l_{\mathrm{e}}|l_{\bar{\mathrm{e}}}\rangle & \Im \langle l_{\mathrm{e}}|l_{\bar{\mathrm{e}}}\rangle \\
\Im \langle l_{\mathrm{e}}|l_{\bar{\mathrm{e}}}\rangle & 1-\Re\langle l_{\mathrm{e}}|l_{\bar{\mathrm{e}}}\rangle
\end{bmatrix}.
\end{eqnarray}
This \emph{real} overlap matrix can be utilized to produce a time-reversal invariant orthonormal basis.
\end{enumerate}

(ii) For the odd-electron case, an orthonormal set of $\{|l_{\mathrm{o}}\rangle\}$ will not be automatically orthogonal to $\{|l_{\bar{\mathrm{o}}}\rangle\}$,
except for $\langle l_{\mathrm{o}}|l_{\bar{\mathrm{o}}}\rangle=0$.
This is different from the nonrelativistic or spin-free relativistic case with $S_z$ symmetry,
where the two parts are of different spin projections and hence are orthogonal automatically.
In general, the overlap metric for $\{|l_{\mathrm{o}}\rangle,\;|l_{\bar{\mathrm{o}}}\rangle\}$ has a quaternion structure
\begin{eqnarray}
\mathbf{S}=
\begin{bmatrix}
\mathbf{A} & \mathbf{B} \\
-\mathbf{B}^* & \mathbf{A}^*
\end{bmatrix},
\quad
\mathbf{A}^\dagger=\mathbf{A},\quad
\mathbf{B}^T=-\mathbf{B}.\label{eq:quaternion}
\end{eqnarray}
Diagonalizing it with an algorithm preserving the quaternion structure\cite{rosch1983time,dongarra1984eigenvalue,bunse1989quaternion,
saue1999quaternion,peng2009construction,shiozaki2017efficient,li2021structured} produce the following
structured eigenvectors
\begin{eqnarray}
\mathbf{Z}=
\begin{bmatrix}
\mathbf{X} & -\mathbf{Y}^* \\
\mathbf{Y} & \mathbf{X}^* \\
\end{bmatrix},\quad
\mathbf{Z}^\dagger\mathbf{Z}=\mathbf{I},\quad
\mathbf{Z}^\dagger\mathbf{S}\mathbf{Z}=
\begin{bmatrix}
\mathbf{\Lambda} & \mathbf{0} \\
\mathbf{0} & \mathbf{\Lambda} \\
\end{bmatrix},
\end{eqnarray}
where $\mathbf{\Lambda}$ is a positive semidefinite diagonal matrix.
The matrix $\mathbf{Z}$ is also symplectic
\begin{eqnarray}
\mathbf{Z}^T\mathbf{J}\mathbf{Z}=
\mathbf{J},\quad
\mathbf{J}=
\begin{bmatrix}
\mathbf{0} & \mathbf{I} \\
-\mathbf{I} & \mathbf{0}
\end{bmatrix}.
\end{eqnarray}
It can be used to construct a time-reversal paired orthonormal basis,
e.g., using canonical orthonormalization\cite{szabo2012modern}.

Therefore, we demonstrate that the basis with the structure \eqref{eq:VlRel} does exist for a time-reversal invariant subspace $V_l$.
In fact, Eq. \eqref{eq:VlRel} corresponds to the only two classes of irreducible projective representations of $U(1)\rtimes Z_2^T$
for systems with particle number $U(1)$ and time-reversal symmetry $Z_2^T$\cite{chen2012symmetry}.
Here, $\rtimes$ represents a semidirect product, because for any element $e^{i\hat{N}\phi}$ in $U(1)$ with
$\hat{N}$ being the particle number operator, we have $\mathcal{T} e^{i\hat{N}\phi} =
e^{-i\hat{N}\phi}\mathcal{T}$ instead of a commuting relation.
If $V_l$ is not invariant under the action of $\mathcal{T}$, we will refer it as time-reversal incomplete.
Calculations performed within such space (e.g., configuration interaction) can be called Kramers
symmetry contaminated, in analogy to spin contamination in the nonrelativistic case.

\subsection{Direct product space}
Having defined the time-reversal symmetry adapted basis, we consider its
construction in a direct product space, which is relevant for
constructing a configuration interaction space in DMRG.
The direct product space of $V_l$ and another subspace $V_r$ with the
structure \eqref{eq:VlRel} can be decomposed as
\begin{eqnarray}
V_l\otimes V_r &=& (V_l^{\mathrm{e}}\oplus V_l^{\mathrm{o}})
\otimes (V_r^{\mathrm{e}}\oplus V_r^{\mathrm{o}}),\nonumber\\
(V_l\otimes V_r)^{\mathrm{e}} &=& (V_l^{\mathrm{e}}\otimes V_r^{\mathrm{e}})\oplus
(V_l^{\mathrm{o}}\otimes V_r^{\mathrm{o}}) \nonumber\\
&=&\spn
\{|l_{\mathrm{e}}r_{\mathrm{e}}\rangle,\;
|l_{\mathrm{o}}r_{\mathrm{o}}\rangle,\;
|l_{\mathrm{o}}r_{\bar{\mathrm{o}}}\rangle,\;
|l_{\bar{\mathrm{o}}}r_{\bar{\mathrm{o}}}\rangle,\;
|l_{\bar{\mathrm{o}}}r_{\mathrm{o}}\rangle\},\nonumber\\
(V_l\otimes V_r)^{\mathrm{o}} &=& (V_l^{\mathrm{e}}\otimes V_r^{\mathrm{o}})\oplus
(V_l^{\mathrm{o}}\otimes V_r^{\mathrm{e}}) \nonumber\\
&=&
\spn
\{|l_{\mathrm{e}}r_{\mathrm{o}}\rangle,\;|l_{\mathrm{o}}r_{\mathrm{e}}\rangle,\;
|l_{\mathrm{e}}r_{\bar{\mathrm{o}}}\rangle,\;|l_{\bar{\mathrm{o}}}r_{\mathrm{e}}\rangle\}.\label{eq:Vlr}
\end{eqnarray}
The pair structure for basis functions of $(V_l\otimes V_r)^{\mathrm{o}}$ is clear,
following from the structures of $V_l$ and $V_r$.
For the even-electron subspace $(V_l\otimes V_r)^{\mathrm{e}}$, by noting that
\begin{eqnarray}
\mathcal{T}|l_{\mathrm{o}}r_{\mathrm{o}}\rangle = |l_{\bar{\mathrm{o}}}r_{\bar{\mathrm{o}}}\rangle,\quad
\mathcal{T}|l_{\mathrm{o}}r_{\bar{\mathrm{o}}}\rangle = -|l_{\bar{\mathrm{o}}} r_{\mathrm{o}}\rangle,
\end{eqnarray}
we can define the following time-reversal invariant basis
\begin{equation}
\left\{
\begin{array}{lll}
|\Phi_{l_{\mathrm{e}}r_{\mathrm{e}}}\rangle &=& |l_{\mathrm{e}}r_{\mathrm{e}}\rangle, \\
|\Phi_{l_{\mathrm{o}}r_{\mathrm{o}}}^{+}\rangle &=& \frac{1}{\sqrt{2}}(|l_{\mathrm{o}}r_{\mathrm{o}}\rangle + |l_{\bar{\mathrm{o}}}r_{\bar{\mathrm{o}}}\rangle), \\
|\Phi_{l_{\mathrm{o}}r_{\mathrm{o}}}^{-}\rangle &=& \frac{\ii}{\sqrt{2}}(|l_{\mathrm{o}}r_{\mathrm{o}}\rangle - |l_{\bar{\mathrm{o}}}r_{\bar{\mathrm{o}}}\rangle), \\
|\Phi_{l_{\mathrm{o}}r_{\bar{\mathrm{o}}}}^{+}\rangle &=& \frac{\ii}{\sqrt{2}}(|l_{\mathrm{o}}r_{\bar{\mathrm{o}}}\rangle + |l_{\bar{\mathrm{o}}}r_{\mathrm{o}}\rangle), \\
|\Phi_{l_{\mathrm{o}}r_{\bar{\mathrm{o}}}}^{-}\rangle &=& \frac{1}{\sqrt{2}}(|l_{\mathrm{o}}r_{\bar{\mathrm{o}}}\rangle - |l_{\bar{\mathrm{o}}}r_{\mathrm{o}}\rangle).
\end{array}\right.\label{eq:TRbasisfunctions}
\end{equation}
Consider the example that both $|l_{\mathrm{o}}\rangle$ and $|r_{\mathrm{o}}\rangle$
are one-electron spin states, then $\{|\Phi_{l_{\mathrm{o}}r_{\mathrm{o}}}^{+}\rangle,|\Phi_{l_{\mathrm{o}}r_{\mathrm{o}}}^{-}\rangle,|\Phi_{l_{\mathrm{o}}r_{\bar{\mathrm{o}}}}^{+}\rangle\}$
are triplets in a Cartesian representation\cite{dyall2007introduction}, while $|\Phi_{l_{\mathrm{o}}r_{\bar{\mathrm{o}}}}^{-}\rangle$
is singlet, viz.,
\begin{eqnarray}
\left\{
\begin{array}{lll}
S_{1x} &=& \frac{1}{\sqrt{2}}(\alpha\alpha + \beta\beta), \\
S_{1y} &=& \frac{\ii}{\sqrt{2}}(\alpha\alpha - \beta\beta), \\
\ii S_{1z} &=& \frac{\ii}{\sqrt{2}}(\alpha\beta + \beta\alpha), \\
S_0 &=& \frac{1}{\sqrt{2}}(\alpha\beta - \beta\alpha), \\
\end{array}\right.
\end{eqnarray}

Suppose a wavefunction $|\Psi\rangle$  is expanded in this direct product basis is $V_l\otimes V_r=\{|lr\rangle\}$,
the coefficients of $|\Psi\rangle$ and its time-reversal partner $|\bar{\Psi}\rangle\triangleq\mathcal{T}|\Psi\rangle$ are related by
\begin{eqnarray}
\langle lr|\bar{\Psi}\rangle = (\mathcal{T}\langle lr|\bar{\Psi}\rangle)^*
= \langle\bar{l}\bar{r}|\bar{\bar{\Psi}}\rangle^*
= \pm\langle\bar{l}\bar{r}|\Psi\rangle^*,\label{eq:lrPsi}
\end{eqnarray}
where the positive sign is for even-electron systems and the minus sign is for odd-electron systems as $\mathcal{T}^2|\Psi\rangle=-|\Psi\rangle$.
We can write the wavefunction coefficient $\Psi_{lr}=\langle lr|\Psi\rangle$ in the direct product space
as a matrix
\begin{eqnarray}
\mathbf{\Psi}
=\begin{bmatrix}
\Psi_{\mathrm{e}\mathrm{e}} & \Psi_{\mathrm{e}\mathrm{o}} & \Psi_{\mathrm{e}\bar{\mathrm{o}}} \\
\Psi_{\mathrm{o}\mathrm{e}} & \Psi_{\mathrm{o}\mathrm{o}} & \Psi_{\mathrm{o}\bar{\mathrm{o}}} \\
\Psi_{\bar{\mathrm{o}}\mathrm{e}} & \Psi_{\bar{\mathrm{o}}\mathrm{o}} & \Psi_{\bar{\mathrm{o}}\bar{\mathrm{o}}} \\
\end{bmatrix},\label{eq:Psi}
\end{eqnarray}
such that for odd-electron systems
\begin{eqnarray}
\mathbf{\Psi}_{\mathrm{o}}
=\begin{bmatrix}
0 & \Psi_{\mathrm{e}\mathrm{o}} & \Psi_{\mathrm{e}\bar{\mathrm{o}}} \\
\Psi_{\mathrm{o}\mathrm{e}} & 0 & 0 \\
\Psi_{\bar{\mathrm{o}}\mathrm{e}} & 0 & 0 \\
\end{bmatrix},
\;\;
\mathbf{\Psi}_{\bar{\mathrm{o}}}
=\begin{bmatrix}
0 & -\Psi^*_{\mathrm{e}\bar{\mathrm{o}}} & \Psi^*_{\mathrm{e}\mathrm{o}} \\
-\Psi^*_{\bar{\mathrm{o}}\mathrm{e}} & 0 & 0 \\
\Psi^*_{\mathrm{o}\mathrm{e}} & 0 & 0 \\
\end{bmatrix},\label{eq:OddElectronPsi}
\end{eqnarray}
and for even-electron systems
\begin{eqnarray}
\mathbf{\Psi}_{\mathrm{e}}
=\begin{bmatrix}
\Psi_{\mathrm{e}\mathrm{e}} & 0 & 0 \\
0 & \Psi_{\mathrm{o}\mathrm{o}} & \Psi_{\mathrm{o}\bar{\mathrm{o}}} \\
0 & \Psi_{\bar{\mathrm{o}}\mathrm{o}} & \Psi_{\bar{\mathrm{o}}\bar{\mathrm{o}}} \\
\end{bmatrix},
\;\;
\mathbf{\Psi}_{\bar{\mathrm{e}}}
=\begin{bmatrix}
\Psi^*_{\mathrm{e}\mathrm{e}} & 0 & 0 \\
0 & \Psi^*_{\bar{\mathrm{o}}\bar{\mathrm{o}}} & -\Psi^*_{\bar{\mathrm{o}}\mathrm{o}} \\
0 & -\Psi^*_{\mathrm{o}\bar{\mathrm{o}}} & \Psi^*_{\mathrm{o}\mathrm{o}} \\
\end{bmatrix}.
\end{eqnarray}
Similar to Eq. \eqref{eq:VlRel}, we can require the many-electron wavefunction
of even electron system to be time-reversal invariant. Consequently, the
wavefunction coefficient matrix is simplified as
\begin{eqnarray}
\mathbf{\Psi}_{\mathrm{e}}
=\begin{bmatrix}
\Psi_{\mathrm{e}\mathrm{e}} & 0 & 0 \\
0 & \Psi_{\mathrm{o}\mathrm{o}} & \Psi_{\mathrm{o}\bar{\mathrm{o}}} \\
0 & -\Psi_{\mathrm{o}\bar{\mathrm{o}}}^* & \Psi_{\mathrm{o}\mathrm{o}}^* \\
\end{bmatrix}
=\mathbf{\Psi}_{\bar{\mathrm{e}}},\label{eq:EvenElectronPsi}
\end{eqnarray}
with the submatrix $\Psi_{\mathrm{e}\mathrm{e}}$ being real.

\section{Time reversal symmetry-adapted DMRG}
We will show how the structure \eqref{eq:VlRel} can be
maintained during the sweep optimization in DMRG,
and used to reduce memory and computational cost.
For this purpose, it suffices to discuss the local
optimization problem in the sweep optimization,
which amounts to first solve a configuration interaction
problem in the direct product space $V_l\otimes V_r$,
\begin{eqnarray}
\hat{H}|\Psi\rangle = E|\Psi\rangle,\quad |\Psi\rangle\in V_l\otimes V_r,\label{eq:localCI}
\end{eqnarray}
and then produce an optimized basis for $V_l$ or $V_r$
(so-called decimation). We refer the readers to
Refs. \cite{chan2002highly} for a detailed description of
the entire sweep optimization.

\subsubsection{Hamiltonian}
We assume the one-electron basis has a Kramers paired structure $\{\psi_p,\psi_{\bar{p}}\}$,
which can either be spin-orbitals or spinors computed from spin-restricted or Kramers-restricted
self-consistent field calculations, respectively. The action of the time-reversal symmetry operator $\mathcal{T}$
on spin-orbitals/spinors reads
\begin{eqnarray}
\mathcal{T} |\psi_p\rangle = |\psi_{\bar{p}}\rangle,\quad
\mathcal{T} |\psi_{\bar{p}}\rangle = -|\psi_p\rangle.
\end{eqnarray}
In the absence of magnetic field, $\mathcal{T}$ commutes with the Hamiltonian $\hat{H}$, which is written
in a second quantized form as
\begin{eqnarray}
\hat{H} = \sum_{pq} h_{pq} a_p^\dagger a_q + \frac{1}{4}\sum_{pqrs}\langle pq\|sr\rangle a_p^\dagger a_q^\dagger a_r a_s.
\end{eqnarray}
To get the representation of $H$ in
the direct product space $V_l\otimes V_r$, $\hat{H}$ is usually rewritten as
\begin{equation}
\begin{array}{rcl}
\hat{H} &=& \hat{H}^{\mathrm{L}}+\hat{H}^{\mathrm{R}} \\
&+& \sum_{p_{\mathrm{L}}} (a_{p_{\mathrm{L}}}^{\mathrm{L}\dagger} S_{p_{\mathrm{L}}}^{\mathrm{R}} + \hc) + \sum_{q_{\mathrm{R}}} (a_{q_{\mathrm{R}}}^{\mathrm{R}\dagger} S_{q_{\mathrm{R}}}^{\mathrm{L}} + \hc) \\
&+& \sum_{p_{\mathrm{L}}<q_{\mathrm{L}}}(A_{p_{\mathrm{L}}q_{\mathrm{L}}}^{\mathrm{L}} P_{p_{\mathrm{L}}q_{\mathrm{L}}}^{\mathrm{R}} + \hc) \\
&+& \sum_{p_{\mathrm{L}}\le s_{\mathrm{L}}}w_{p_{\mathrm{L}}s_{\mathrm{L}}}(B_{p_{\mathrm{L}}s_{\mathrm{L}}}^{\mathrm{L}} Q_{p_{\mathrm{L}}s_{\mathrm{L}}}^{\mathrm{R}} + \hc),\label{eq:HlRdecomp}
\end{array}
\end{equation}
where $w_{pq}=1-\frac{1}{2}\delta_{pq} = w_{qp}$, $p_{\mathrm{L}}$ ($p_{\mathrm{R}}$) represents the index of one-electron basis for the subspace $V_l$ ($V_r$),
and the introduced intermediates (normal and complementary operators\cite{xiang1996density,chan2002highly}) are
\begin{equation}
\begin{array}{rcl}
A_{pq} &\triangleq& a_{p}^\dagger a_{q}^\dagger, \\
B_{ps} &\triangleq& a_{p}^\dagger a_{s}, \\
P_{pq} &\triangleq& \sum_{s<r}\langle pq\|sr\rangle a_{r} a_{s}, \\
Q_{ps} &\triangleq& \sum_{qr}\langle pq\|sr\rangle a_{q}^\dagger a_{r}, \\
S_{p} &\triangleq& \sum_q\frac{1}{2}h_{pq}a_{q} + \sum_{q,s<r}\langle pq\|sr\rangle a_{q}^\dagger a_{r} a_{s}.
\end{array}
\end{equation}
The time-reversal invariance of the Hamiltonian $\hat{H}=\mathcal{T}\hat{H}\mathcal{T}^{-1}$ is not obvious
in this form. Since the integrals satisfy time-reversal symmetry, viz.,
$h_{pq}^*=h_{\bar{p}\bar{q}}$ and $h_{p\bar{q}}^*=-h_{\bar{p}q}$,
we can rewrite Eq. \eqref{eq:HlRdecomp} as
\begin{eqnarray}
\begin{array}{rcl}
\hat{H} &=& \tilde{H} + \mathcal{T}\tilde{H}\mathcal{T}^{-1}, \\
\tilde{H} &\triangleq& \frac{1}{2}(\hat{H}^{\mathrm{L}} + \hat{H}^{\mathrm{R}}) + (H_{aS} + H_{AP} + H_{BQ}), \\
H_{aS}&\triangleq& \sum_{p_{\mathrm{L}}}a_{p_{\mathrm{L}}}^{\mathrm{L}\dagger} S_{p_{\mathrm{L}}}^{\mathrm{R}} + \sum_{p_{\mathrm{R}}}a_{p_{\mathrm{R}}}^{\mathrm{R}\dagger} S_{p_{\mathrm{R}}}^{\mathrm{L}} + \hc, \\
H_{AP}&\triangleq& \sum_{p_{\mathrm{L}}<q_{\mathrm{L}}} A^{\mathrm{L}}_{p_{\mathrm{L}}q_{\mathrm{L}}} P^{\mathrm{R}}_{p_{\mathrm{L}}q_{\mathrm{L}}} \\
&&+\sum_{p_{\mathrm{L}}\le q_{\mathrm{L}}} w_{p_{\mathrm{L}}q_{\mathrm{L}}} A^{\mathrm{L}}_{p_{\mathrm{L}}\bar{q}_{\mathrm{L}}} P^{\mathrm{R}}_{p_{\mathrm{L}}\bar{q}_{\mathrm{L}}} + \hc, \\
H_{BQ}&\triangleq& \sum_{p_{\mathrm{L}}\le s_{\mathrm{L}}} w_{p_{\mathrm{L}}s_{\mathrm{L}}} B^{\mathrm{L}}_{p_{\mathrm{L}}s_{\mathrm{L}}} Q^{\mathrm{R}}_{p_{\mathrm{L}}s_{\mathrm{L}}} \\
&&+\sum_{p_{\mathrm{L}}\le s_{\mathrm{L}}} w_{p_{\mathrm{L}}s_{\mathrm{L}}} B^{\mathrm{L}}_{p_{\mathrm{L}}\bar{s}_{\mathrm{L}}} Q^{\mathrm{R}}_{p_{\mathrm{L}}\bar{s}_{\mathrm{L}}} + \hc,
\end{array}\label{eq:HamKRS}
\end{eqnarray}
using the derived time-reversal symmetry properties of intermediates such as
\begin{eqnarray}
\mathcal{T} A_{pq} \mathcal{T}^{-1} &=& \mathcal{T} (p^\dagger q^\dagger) \mathcal{T}^{-1}
= \bar{p}^\dagger \bar{q}^\dagger = A_{\bar{p}\bar{q}}, \\
\mathcal{T} A_{p\bar{q}} \mathcal{T}^{-1} &=& \mathcal{T} (p^\dagger \bar{q}^\dagger) \mathcal{T}^{-1}
= -\bar{p}^\dagger q^\dagger = -A_{\bar{p}q}.
\end{eqnarray}
The obtained 'skeleton' operator $\tilde{H}$ is Hermitian but not time-reversal invariant,
but the number of operators in $\tilde{H}$ is roughly half of
that in $\hat{H}$. This form will be used later to
reduce the computational cost of R-DMRG.

\subsubsection{Diagonalization}
To solve Eq. \eqref{eq:localCI} in $V_l\otimes V_r$ with $\hat{H}$ in Eq. \eqref{eq:HamKRS},
we can use the iterative Davidson algorithm, where in each step the so-called
$\sigma$ vector needs to be formed $|\sigma\rangle=\hat{H}|\Psi\rangle$.
For the even-electron case, the coefficients of $\sigma$ are
\begin{eqnarray}
\sigma_{lr}^{\mathrm{e}} &\triangleq& \langle lr|\tilde{H}+\mathcal{T}\tilde{H}\mathcal{T}^{-1}|\Psi_{\mathrm{e}}\rangle \nonumber\\
&=& \langle lr|\tilde{H}|\Psi_{\mathrm{e}}\rangle + \langle\bar{l}\bar{r}|\tilde{H}|\Psi_{\bar{\mathrm{e}}}\rangle^*.\label{eq:TRSymmetrization0}
\end{eqnarray}
Since we require $|\Psi_{\mathrm{e}}\rangle=|\Psi_{\bar{\mathrm{e}}}\rangle$, it simply becomes
\begin{eqnarray}
\sigma_{lr}^{\mathrm{e}} = \tilde{\sigma}_{lr}^{\mathrm{e}} + \tilde{\sigma}_{\bar{l}\bar{r}}^{\mathrm{e}*},\label{eq:TRSymmetrization}
\end{eqnarray}
with $\tilde{\sigma}_{lr}^{\mathrm{e}}\triangleq \langle lr|\tilde{H}|\Psi_{\mathrm{e}}\rangle$
This shows that only $\tilde{\sigma}_{lr}^{\mathrm{e}}$ needs to be constructed,
whose computational cost is roughly half of that for constructing $\sigma_{lr}^{\mathrm{e}}$.
Similarly, for odd-electron systems, we can find
\begin{eqnarray}
\sigma_{lr}^{\mathrm{o}} &\triangleq& \langle lr|\tilde{H}+\mathcal{T}\tilde{H}\mathcal{T}^{-1}|\Psi_{\mathrm{o}}\rangle \nonumber\\
&=& \langle lr|\tilde{H}|\Psi^{\mathrm{o}}\rangle + \langle\bar{l}\bar{r}|\tilde{H}|\Psi_{\bar{\mathrm{o}}}\rangle^*
= \tilde{\sigma}_{lr}^{\mathrm{o}} + \tilde{\sigma}_{\bar{l}\bar{r}}^{\bar{\mathrm{o}}*}, \\
\sigma_{lr}^{\bar{\mathrm{o}}} &\triangleq& \langle lr|\tilde{H}+\mathcal{T}\tilde{H}\mathcal{T}^{-1}|\Psi_{\bar{\mathrm{o}}}\rangle \nonumber\\
&=& \langle lr|\tilde{H}|\Psi_{\bar{\mathrm{o}}}\rangle - \langle\bar{l}\bar{r}|\tilde{H}|\Psi_{\mathrm{o}}\rangle^*
= \tilde{\sigma}_{lr}^{\bar{\mathrm{o}}} - \tilde{\sigma}_{\bar{l}\bar{r}}^{\mathrm{o}*}.
\end{eqnarray}
Thus, it suffices to construct $\tilde{\sigma}_{lr}^{\mathrm{o}}$ and $\tilde{\sigma}_{lr}^{\bar{\mathrm{o}}}$,
which reduces the computational cost for constructing $\sigma_{lr}^{\mathrm{o}}$ and $\sigma_{lr}^{\bar{\mathrm{o}}}$ roughly
by half.

In summary, the full $\sigma$ can be recovered from the skeleton one $\tilde{\sigma}$
by a 'time-reversal symmetrization' in both even- and odd-electron cases. This is in
a similar spirit to the construction of Fock matrix using the
skeleton-matrix algorithm\cite{dacre1970use,elder1973use,dupuis1977molecular}.
Expressions for $\tilde{H}$ in Eq. \eqref{eq:HamKRS}
immediately show that only half of the intermediate operators are necessary for
constructing $\tilde{H}$, which reduces the memory and computational cost for renormalized operators by half
compared with an implementation without using time-reversal symmetry.

To use such reduction, we need to maintain the structure \eqref{eq:VlRel}
for basis vectors of the subspace in Davidson algorithm.
For the even-electron case, suppose the current subspace $V=\spn\{|b_k\rangle\}$ in Davidson algorithm
is spanned by time-reversal invariant basis, then the representation of $\hat{H}$ is a real matrix,
\begin{eqnarray}
\langle b_k|H|b_l\rangle = (\mathcal{T}\langle b_k|H|b_l\rangle)^* =
\langle \bar{b}_k|H|\bar{b}_l\rangle = \langle b_k|H|b_l\rangle.
\end{eqnarray}
Consequently, the eigenvectors $\mathbf{X}$ of $\mathbf{H}$ are real and
the states $|x_i\rangle= \sum_{k}|b_k\rangle X_{ki}$ are time-reversal invariant.
It can be verified that the residual $|r_i\rangle=\hat{H}|x_i\rangle - |x_i\rangle E_i$ is also time-reversal invariant,
so does the precondition residual as $\langle lr|H|lr\rangle = \langle\bar{l}\bar{r}|H|\bar{l}\bar{r}\rangle$.

For the odd-electron case, suppose the current subspace $V=\spn\{|b_k\rangle\}\oplus\spn\{|\bar{b}_k\rangle\}$ is spanned
by a Kramers paired basis, i.e., $|\bar{b}_k\rangle = \mathcal{T}|b_k\rangle$,
then the representation of Hamiltonian has a quaternion structure
\eqref{eq:quaternion}. Diagonalizing it with a structure-preserving
algorithm will produced Kramers paired eigenvectors $\{|x_i\rangle,|\bar{x}_i\rangle\}$.
Similarly, one can show that the residuals also form a Kramers pair,
$|\bar{r}_i\rangle = \mathcal{T}|r_i\rangle$. An important point
for constructing new Kramers paired orthonormal basis is that
if $|r_i\rangle$ is already orthonormalized against $V=\spn\{|b_k\rangle\}\oplus\spn\{|\bar{b}_k\rangle\}$,
then $|\bar{r}_i\rangle$ will be automatically orthogonal to the basis vectors
in $V$, such that the pair $(|r_i\rangle,|\bar{r}_i\rangle)$ can be added
simultaneously. This property can be seen from
\begin{eqnarray}
\langle\bar{r}_i|b_k\rangle &=& (\mathcal{T}\langle\bar{r}_i|b_k\rangle)^* = -\langle r_i|\bar{b}_k\rangle = 0,\\
\langle\bar{r}_i|\bar{b}_k\rangle &=& (\mathcal{T}\langle\bar{r}_i|\bar{b}_k\rangle)^* = \langle r_i|b_k\rangle = 0,
\end{eqnarray}
and $\langle\bar{r}_i|r_i\rangle=0$ due to Kramers' theorem. In this
way, we can maintain the structure \eqref{eq:VlRel} for the basis vectors
of the subspace in the Davidson diagonalization algorithm.

\subsubsection{Decimation}
Once the eigenvectors $|\Psi\rangle$ \eqref{eq:Psi} of $\hat{H}$ have been found in the direct product space $V_l\otimes V_r$,
we need to perform decimation to obtain an optimized basis for $V_l$ or $V_r$.
This is done by diagonalizing the reduced density matrix $\rho_l$
or $\rho_r$. In the sweep optimization of DMRG, $V_l$ (or $V_r$) is also a direct
product space denoted by $V_D$. It is formed by a direct product between the left environment
and the left dot, referred as superblock. Simply diagonalizing the reduced density
matrix will not yield a basis with the structure \eqref{eq:VlRel}.
We will show how to perform decimation in such space to produce time-reversal
symmetry-adapted renormalized states.

For the even-electron subspace $V_D^{\mathrm{e}}$, we assume it is spanned by
the following direct product basis
\begin{eqnarray}
V_D^{\mathrm{e}} = \spn\{|D_{\mathrm{e}}\rangle,|D_{\bar{\mathrm{e}}}\rangle,|D_0\rangle\},\label{eq:VDe}
\end{eqnarray}
with $|D_{\bar{\mathrm{e}}}\rangle=\mathcal{T}|D_{\mathrm{e}}\rangle$ and $|D_0\rangle = \mathcal{T}|D_0\rangle$.
Here, $|D_0\rangle$ represents the part of direct product basis which
is already time-reversal invariant such as $|l_{\mathrm{e}}r_{\mathrm{e}}\rangle$ in Eq. \eqref{eq:Vlr}.
The pair $|D_{\mathrm{e}}\rangle$ and $|D_{\bar{\mathrm{e}}}\rangle$ represent those parts
which can be related by $\mathcal{T}$ such as $|l_{\mathrm{o}}r_{\mathrm{o}}\rangle$ and $|l_{\bar{\mathrm{o}}}\bar{r}_{\mathrm{o}}\rangle$.
Suppose the reduced density matrix obtained from $|\Psi\rangle$ in $V_D$ is
\begin{eqnarray}
\boldsymbol{\rho}^\Psi=\left[\begin{array}{ccc}
\rho_{\mathrm{e}\mathrm{e}} & \rho_{e\bar{\mathrm{e}}} & \rho_{e0} \\
\rho_{\bar{\mathrm{e}}e} & \rho_{\bar{\mathrm{e}}\bar{\mathrm{e}}} & \rho_{\bar{\mathrm{e}}0} \\
\rho_{0e} & \rho_{0\bar{\mathrm{e}}} & \rho_{00} \\
\end{array}\right],
\end{eqnarray}
then that obtained from $|\bar{\Psi}\rangle=\mathcal{T}|\Psi\rangle$ is
\begin{eqnarray}
\boldsymbol{\rho}^{\bar{\Psi}}=\left[\begin{array}{ccc}
\rho^*_{\bar{\mathrm{e}}\bar{\mathrm{e}}} & \rho^*_{\bar{\mathrm{e}}e} & \rho^*_{\bar{\mathrm{e}}0} \\
\rho^*_{e\bar{\mathrm{e}}} & \rho^*_{\mathrm{e}\mathrm{e}} & \rho^*_{e0} \\
\rho^*_{0\bar{\mathrm{e}}} & \rho^*_{0e} & \rho^*_{00} \\
\end{array}\right].
\end{eqnarray}
The average $\boldsymbol{\rho}=\frac{1}{2}(\boldsymbol{\rho}^\Psi+\boldsymbol{\rho}^{\bar{\Psi}})$ is time-reversal invariant
\begin{eqnarray}
\boldsymbol{\rho} =\frac{1}{2}\left[\begin{array}{ccc}
\rho_{\mathrm{e}\mathrm{e}}+\rho^*_{\bar{\mathrm{e}}\bar{\mathrm{e}}} & \rho_{e\bar{\mathrm{e}}}+\rho^*_{\bar{\mathrm{e}}e} & \rho_{e0}+\rho^*_{\bar{\mathrm{e}}0} \\
\rho_{\bar{\mathrm{e}}e}+\rho^*_{e\bar{\mathrm{e}}} & \rho_{\bar{\mathrm{e}}\bar{\mathrm{e}}}+\rho^*_{\mathrm{e}\mathrm{e}} & \rho_{\bar{\mathrm{e}}0}+\rho^*_{e0} \\
\rho_{0e}+\rho^*_{0\bar{\mathrm{e}}} & \rho_{0\bar{\mathrm{e}}}+\rho^*_{0e} & \rho_{00}+\rho^*_{00} \\
\end{array}\right]
\triangleq
\left[\begin{array}{ccc}
\mathcal{A} & \mathcal{B} & \mathcal{C} \\
\mathcal{B}^* & \mathcal{A}^* & \mathcal{C}^* \\
\mathcal{C}^\dagger & \mathcal{C}^T & \mathcal{E} \\
\end{array}\right],\label{eq:RrhoAVeven0}
\end{eqnarray}
where $\mathcal{A}=\mathcal{A}^\dagger$, $\mathcal{B}=\mathcal{B}^T$, and $\mathcal{E}$ is
real symmetric. However, simply diagonalizing it with a complex eigensolver will
not produce time-reversal invariant basis function \eqref{eq:VlRel}
due to the arbitrariness of the phase factor.
To fix this problem, we can introduce a time-reversal invariant basis similar to Eq. \eqref{eq:TRbasisfunctions},
\begin{eqnarray}
(|R_-\rangle,|R_+\rangle,|R_0\rangle) \triangleq (|R_{\mathrm{e}}\rangle,|R_{\bar{\mathrm{e}}}\rangle,|R_0\rangle)\mathbf{U}
\end{eqnarray}
with $\mathbf{U}$ being
\begin{eqnarray}
\mathbf{U} = \left[
\begin{array}{ccc}
\frac{\ii}{\sqrt{2}}I & \frac{1}{\sqrt{2}}I & 0 \\
-\frac{\ii}{\sqrt{2}}I & \frac{1}{\sqrt{2}}I & 0 \\
0 & 0 & I \\
\end{array}\right],
\end{eqnarray}
and transform $\boldsymbol{\rho}$ into this basis, which leads to a real symmetric
reduced density matrix
\begin{eqnarray}
\tilde{\boldsymbol{\rho}} =
\mathbf{U}^\dagger \boldsymbol{\rho}\mathbf{U}
=
\left[\begin{array}{ccc}
(\mathcal{A}-\mathcal{B})_\rR  & (\mathcal{A}+\mathcal{B})_\rI & \sqrt{2}\mathcal{C}_\rI \\
-(\mathcal{A}-\mathcal{B})_\rI & (\mathcal{A}+\mathcal{B})_\rR & \sqrt{2}\mathcal{C}_\rR \\
\sqrt{2}\mathcal{C}_\rI^T & \sqrt{2}\mathcal{C}_\rR^T & \mathcal{E} \\
\end{array}\right],\label{eq:RrhoAVeven1}
\end{eqnarray}
where $\mathcal{A}_{\rR}$ (or $\mathcal{A}_{\rI}$) represents the real (or imaginary)
part. Diagonalizing $\tilde{\boldsymbol{\rho}}\mathbf{X}=\mathbf{X}\mathbf{\Lambda}$ yields
a set of real vectors $\mathbf{X}$ in the time-reversal invariant basis, which
can be back transformed to the original direct product basis \eqref{eq:VDe} by
\begin{eqnarray}
\mathbf{UX} =
\left[\begin{array}{ccc}
\frac{\ii}{\sqrt{2}}I & \frac{1}{\sqrt{2}}I & 0 \\
-\frac{\ii}{\sqrt{2}}I & \frac{1}{\sqrt{2}}I & 0 \\
0 & 0 & 1 \\
\end{array}\right]
\left[\begin{array}{c}
\mathbf{X}_- \\
\mathbf{X}_+ \\
\mathbf{X}_0 \\
\end{array}\right]
\equiv
\left[\begin{array}{c}
\mathbf{X}_{\mathrm{e}} \\
\mathbf{X}^*_{\mathrm{e}} \\
\mathbf{X}_0 \\
\end{array}\right],
\end{eqnarray}
with $\mathbf{X}_{\mathrm{e}}=(\mathbf{X}_+ +\ii\mathbf{X}_-)/\sqrt{2}$.

For the odd-electron subspace $V_D^{\mathrm{o}}$, we assume it is spanned by
the following direct product basis
\begin{eqnarray}
V_D^{\mathrm{o}} = \spn\{|D_{\mathrm{o}}\rangle,|D_{\bar{\mathrm{o}}}\rangle\},\label{eq:VDo}
\end{eqnarray}
which is already Kramers paired, see Eq. \eqref{eq:Vlr}.
The reduced density matrices are
\begin{eqnarray}
\boldsymbol{\rho}^\Psi = \left[\begin{array}{cc}
\rho_{\mathrm{o}\mathrm{o}} & \rho_{\mathrm{o}\bar{\mathrm{o}}} \\
\rho_{\bar{\mathrm{o}}\mathrm{o}} & \rho_{\bar{\mathrm{o}}\bar{\mathrm{o}}} \\
\end{array}\right],\quad
\boldsymbol{\rho}^{\bar{\Psi}} = \left[\begin{array}{cc}
\rho^*_{\bar{\mathrm{o}}\bar{\mathrm{o}}} & -\rho^*_{\bar{\mathrm{o}}\mathrm{o}} \\
-\rho^*_{\mathrm{o}\bar{\mathrm{o}}} & \rho^*_{\mathrm{o}\mathrm{o}} \\
\end{array}\right],
\end{eqnarray}
and the average $\boldsymbol{\rho}$ has a quaternion structure
\begin{eqnarray}
\boldsymbol{\rho} = \frac{1}{2}
\left[\begin{array}{cc}
\rho_{\mathrm{o}\mathrm{o}}+\rho^*_{\bar{\mathrm{o}}\bar{\mathrm{o}}} & \rho_{\mathrm{o}\bar{\mathrm{o}}}-\rho^*_{\bar{\mathrm{o}}\mathrm{o}} \\
\rho_{\bar{\mathrm{o}}\mathrm{o}}-\rho^*_{\mathrm{o}\bar{\mathrm{o}}}  & \rho_{\bar{\mathrm{o}}\bar{\mathrm{o}}}+\rho^*_{\mathrm{o}\mathrm{o}} \\
\end{array}\right]
\triangleq
\left[\begin{array}{cc}
\mathcal{A} & \mathcal{B} \\
-\mathcal{B}^*  & \mathcal{A}^* \\
\end{array}\right].\label{eq:RrhoAVodd}
\end{eqnarray}
Thus, diagonalizing it with a structural preserving algorithm\cite{rosch1983time,dongarra1984eigenvalue,bunse1989quaternion,
saue1999quaternion,peng2009construction,shiozaki2017efficient,li2021structured} will lead to a new set of Kramers paired renormalized states.

The above decimation procedure is quite general. We can even apply the decimation procedure for cases where $|\Psi\rangle$ is Kramers symmetry contaminated
to produce a time-reversal symmetry-adapted basis. For instance, it can be applied to convert a Kramers symmetry contaminated selected configuration 
interaction wavefunctions to time-reversal symmetry-adapted MPS
as the initial guess for R-DMRG\cite{li2021expressibility}. Then, by iterating the above diagonalization and decimation procedure,
the time-reversal symmetry structure of basis functions \eqref{eq:VlRel} can be maintained recursively during the sweep optimization in DMRG.
For other tensor network states without loops\cite{shi2006classical,murg2010simulating,
nakatani2013efficient,murg2015tree,chepiga2019comb,li2021expressibility},
it is clear that the construction of time-reversal symmetry-adapted basis
can be directly applied.

%\section{Results}
%We implemented time-reversal symmetry adaptation in our
%in-house program \textsc{Focus}\cite{li2021expressibility},
%developed for exploring tensor network algorithms
%for quantum chemistry. We generate the initial time-reversal symmetry-adapted MPS from a crude selected configuration interaction
%calculation using a conversion procedure from Ref. \cite{li2021structured} with the above
%decimation incorporating time-reversal symmetry.
%
%{\color{red}
%H6 and H6+?
%1. illustrate the Kramers symmetry breaking for H5? increasing D
%2. increasing no. of matrix-vector multiplications
%how to compare with the NR case?
%}

\section{Conclusion}
In this work, we propose the time-reversal symmetry adaption for R-DMRG
by introducing a time-reversal symmetry-adapted basis \eqref{eq:VlRel}
and strategies to maintaining this structure during the sweep
optimization in DMRG. It overcomes the artificial symmetry
breaking in conventional R-DMRG calculations and leads to
a reduction of memory and computational cost.
The construction of time-reversal symmetry-adapted basis
also directly applies to other tensor network states without loops.
This opens up new possibilities of applying R-DMRG
for complex heavy-element compounds. Applications of the introduced
time-reversal symmetry-adapted R-DMRG
will be reported in due time.

\section*{Acknowledgements}
This work was supported by the National Natural Science Foundation of China (Grants
No. 21973003) and the Beijing Normal University Startup Package.

\bibliographystyle{apsrev4-1}
\bibliography{references}

\end{document}